\documentclass[fleqn,12pt,twoside]{article}
\usepackage{espcrc1}

\usepackage{graphicx}

\usepackage{bm}% bold math

\newcommand{\vt}{\vec \tau}

\title{Short-range Three-Nucleon Forces Effects on Nucleon-Deuteron Scattering}

\author{S. Ishikawa\thanks{Numerical calculations in this research were supported by Research Center for Computing and Multimedia Studies, Hosei University, under Project No. lab0003.}
\address{Science Research Center, Hosei University, 2-17-1 Fujimi, Chiyoda, Tokyo 102-8160, Japan}
\address{Triangle Universities Nuclear Laboratory, Durham, North Carolina 27708-0308, U.S.A.} 
}
      
\begin{document}

% typeset front matter
\maketitle

\begin{abstract}
Effects of three-nucleon forces arising from the exchange of a pion and a scalar-isoscalar object among three nucleons on nucleon-deuteron scattering observables are studied.
\end{abstract}

%%%%%%%%%%%%%%%%%%%%%%
\section{INTRODUCTION}
%%%%%%%%%%%%%%%%%%%%%%

\begin{figure}[tbh]
\begin{minipage}[t]{75mm}
\includegraphics[scale=0.18]{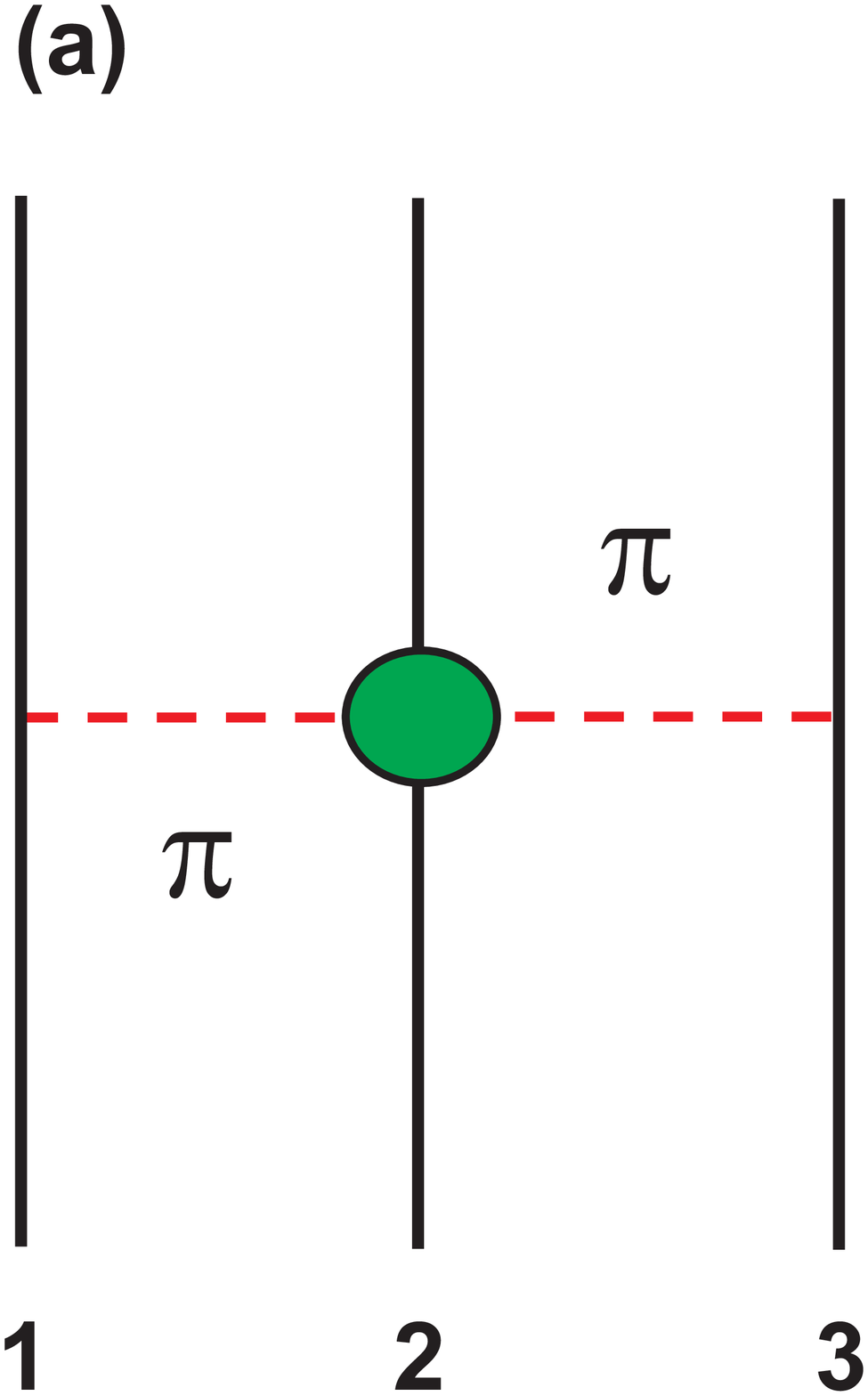}
\includegraphics[scale=0.18]{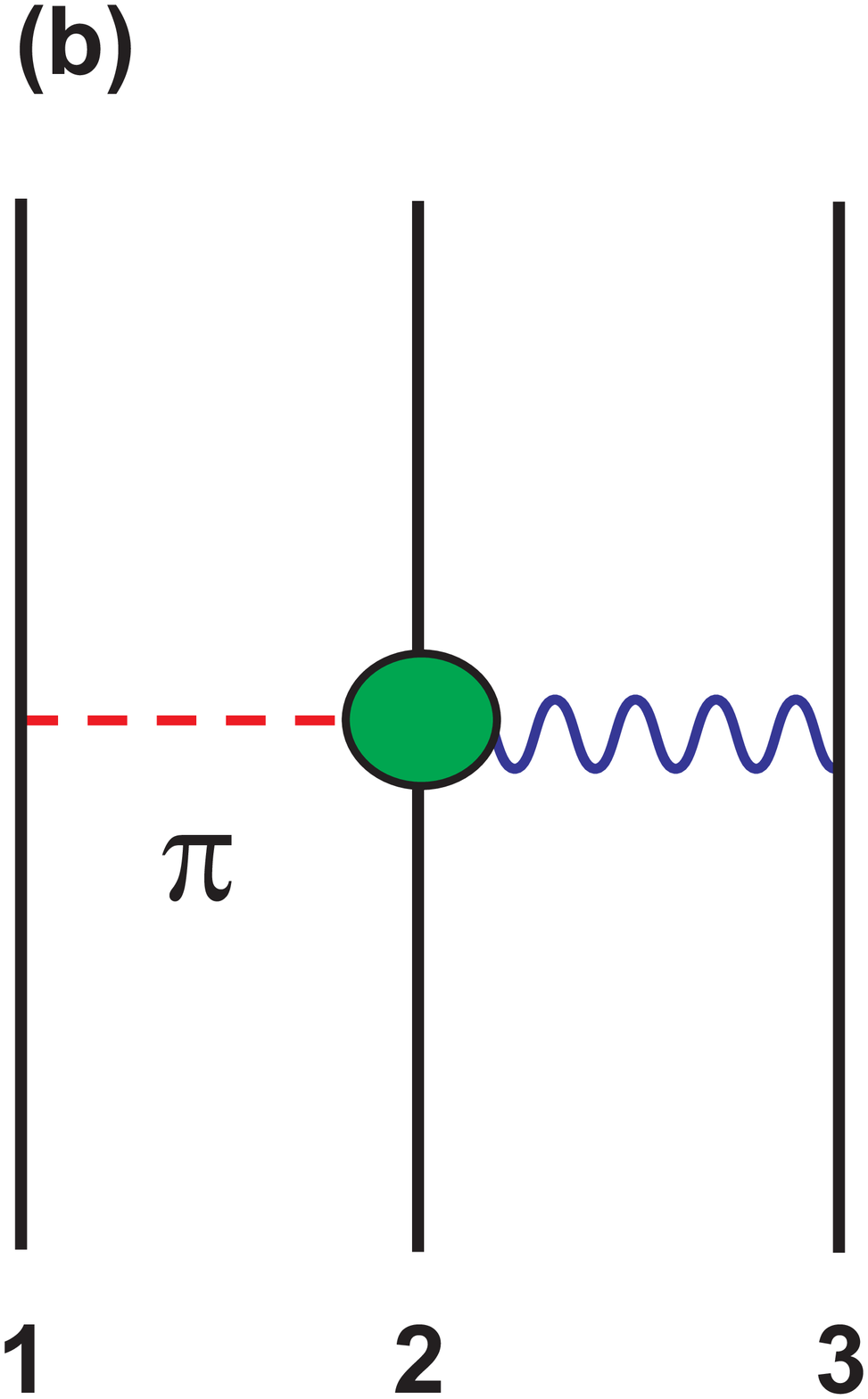}
\caption{
Diagrams for the 2$\pi$E 3NF (a) and the $\pi$-S 3NF (b). 
The wavy line between the nucleons 2 and 3 represents the exchange of a scalar-isoscalar object.
\label{fig:3nf-graph}
}
\end{minipage}
\hspace{\fill}
\begin{minipage}[t]{81mm}
\includegraphics[scale=0.67]{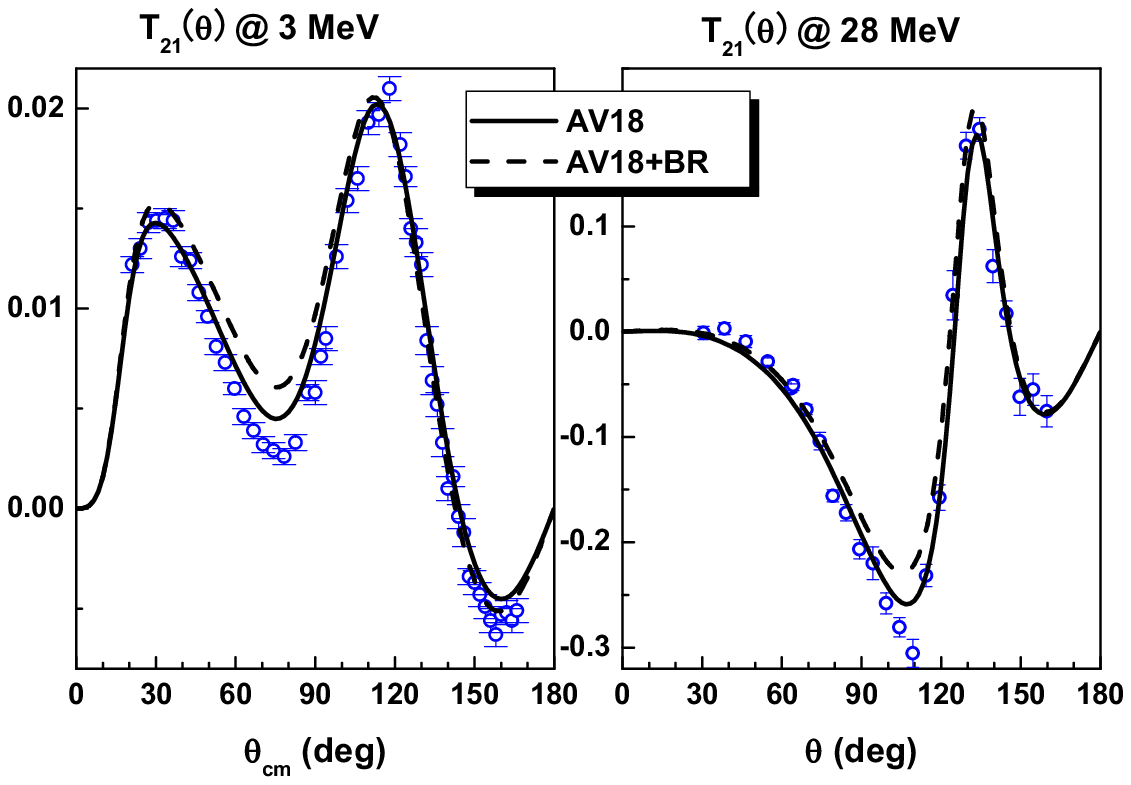}
\caption{ 
$T_{21}(\theta)$ of $pd$ elastic scattering at $E = 3$ MeV and $E = 28$ MeV.
The data are taken from Refs.\ \protect\cite{Sa94,Sh95} for 3 MeV and from Ref.\ \protect\cite{Ha84} for 28 MeV.
\label{fig:t21-br-3-28MeV}
}
\end{minipage}
\end{figure}

The introduction of a three-nucleon force (3NF) arising from the exchange of two pions among three nucleons (2$\pi$E) as shown in Fig.\ \ref{fig:3nf-graph} (a) into nuclear Hamiltonian is known to get rid of discrepancies between experimental data and theoretical calculations with realistic two-nucleon forces (2NFs) for the three-nucleon (3N) binding energies and nucleon-deuteron (ND) differential cross sections. 
On the other hand, the 2$\pi$E 3NF is unsuccessful in explaining some ND polarization observables, {\em e.g.}, too small effects to vector analyzing powers (VAPs) and undesirable contributions to tensor analyzing powers (TAPs) in low-energy ND elastic scattering (see Fig. \ref{fig:t21-br-3-28MeV}). 

In Refs.\ \cite{Is03,Is04}, we have pointed out that tensor components in the 2$\pi$E 3NF should be responsible to the problem in TAPs. 
In this paper, we examine a 3NF due to the exchange of a pion and a scalar-isoscalar object ($\pi$-S) shown in Fig.\ \ref{fig:3nf-graph} (b) as a possible source of tensor interactions that have different characteristics from those of the 2$\pi$E 3NF. 

After giving a general form of the $\pi$-S 3NF in Sec.\ \ref{sec:ps-3nf}, numerical results for the 3N binding energy and ND scattering observables will be presented in Sec.\ \ref{sec:results}.
Summary is given in Sec. \ref{sec:summary}.

%----- FIGURE  -------------------------------------------------------
%---------------------------------------------------------------------

%%%%%%%%%%%%%%%
\section{PION-"SCALAR-ISOSCALAR-OBJECT" EXCHANGE THREE-NUCLEON FORCES}
\label{sec:ps-3nf}
%%%%%%%%%%%%%

We will consider the following models for the $\pi$-S 3NFs:
$\pi$-$\sigma$ exchange with the excitation of the Roper resonance $N^*(1440)$ ($(\pi$-$\sigma)_{N^*}$) \cite{Co95,Ad04}; 
$\pi$-$\sigma$ exchange corresponding to the nucleon Born diagrams (so called pair or Z diagrams) for the PV ($(\pi$-$\sigma)_{Z,PV}$) or the PS ($(\pi$-$\sigma)_{Z,PS}$) $\pi NN$ coupling \cite{Co95,Ad04}; 
$\pi$-"effective scalar field" exchange by a linear model \cite{Ma98} or by a nonlinear model \cite{Ma98}; 
$\pi$-"effective 2$\pi$" exchange ($(\pi$-$2\pi)$) \cite{Ma00}.
A coordinate space representation of these potentials corresponding to diagram  Fig.\ \ref{fig:3nf-graph} (b) is: 
\begin{equation}
W_{12,3}\left(\bm{r}_{12},\bm{r}_{32} \right) =  (\vt_1 \cdot \vt_2)
 (\bm{\sigma}_1\cdot\bm{\nabla}_{12}) 
 \Bigl[  V_a
(\bm{\sigma}_2\cdot\bm{\nabla}_{12}) 
    +  V_b 
 (\bm{\sigma}_2\cdot\bm{\nabla}_{32}) 
 \Bigr]  Y_\pi(r_{12}) Y_\sigma(r_{32}),
\label{eq:Wps}
\end{equation}
where $\bm{r}_{ij} =  \bm{r}_{j}-  \bm{r}_{i}$ 
with $\bm{r}_k$ being a position vector of nucleon $k$.
When a dipole $xNN$ form factor with a cutoff mass $\Lambda_x$ is used, the function $Y_{x}(r)$ becomes
$Y_{x}(r) = e^{- m_x r}/r  
  - \{ 1+  (\Lambda_x^2-m_x^2)r/(2\Lambda_x) \}
e^{- \Lambda_x r}/r$.
Expressions of $V_a$ and $V_b$ for the $\pi$-S 3NFs \cite{Co95,Ad04,Ma98,Ma00} are shown in  Table \ref{tab:pot_strength}.

%%%% Table %%%%%%%%%%%%%%%%%%%%%%%%%%%%%%%%%%%%%%%%%%%%%%%%
\begin{table}[tbh]
\caption{\label{tab:pot_strength} 
The potential strengths $V_a$ and $V_b$ for the $\pi$-S 3NF models and  
and effects of the 3NFs on the ${}^3$H energy: $\Delta E_{3NF} \equiv E_{3,\rm{AV18+3NF}}-E_{3,\rm{AV18}}$. 
}
\begin{tabular}{lccccc}
\hline
  & $V_a$ & \qquad & $V_b$ & \qquad & $\Delta E_{3NF}$ (MeV) \\
\hline
$(\pi$-$\sigma)_{N^*}$ & 
$\frac{g_\pi g_\pi^{*}}{4\pi} \frac{g_\sigma g_\sigma^{*}}{4\pi} \frac{m_\pi^4 }{2(m^*- m_N) m_N^2}$
 & & -- & & -0.32
\\
$(\pi$-$\sigma)_{Z,PV}$ & -- & & 
$-\frac{g_\pi^2}{4\pi} \frac{g_\sigma^2}{4\pi}  \frac{m_\pi^4 }{4m_N^3}$ & & +0.47
\\
$(\pi$-$\sigma)_{Z,PS}$ & 
$\frac{g_\pi^2}{4\pi} \frac{g_\sigma^2}{4\pi} \frac{m_\pi^4 }{4m_N^3}$ 
& & -- & & -1.04
\\
Linear  & 
$\left( \frac{g^2}{4\pi}\right)^2 \frac{m_\pi^4}{4 m_N^3}$ 
& &-- & & -1.99
\\
Nonlinear  & 
$\left( \frac{g^2}{4\pi}\right)^2 \frac{m_\pi^4}{4 m_N^3} \left(1-\frac{m_N}{g f_\pi} \right)$ 
& & 
$-\left( \frac{g^2}{4\pi}\right)^2 \frac{m_\pi^4}{4 m_N^3}\left(\frac{m_N}{g f_\pi} \right)$ & & +0.23
\\
$\pi$-2$\pi$ & 
$- \frac{g_\pi^2}{4\pi} \frac{g_s^2}{4\pi}\frac{m_\pi^4}{4m_N^3}  \frac{m_N m_\pi}{3\alpha_{00}^+ f_\pi^2}$ 
& &
$-\frac{g_\pi^2}{4\pi} \frac{g_s^2}{4\pi} \frac{m_\pi^4}{8m_N^3} \left(\frac1{3\alpha_{00}^+}\frac{m_N m_\pi}{f_\pi^2} +1 \right)$  & & +0.74 
\\
\hline
\end{tabular}
\end{table}
%%%%%%%%%%%%%%%%%%%%%%%%%%%%%%%%%%%%%%%%%%%%%%%%%%%

%%%%%%%%%%%%%%%%%%%%%%%%%%%%%%%%%%%%%%%
\section{NUMERICAL RESULTS}
\label{sec:results}
%%%%%%%%%%%%%%%%%%%%%%%%%%%%%%%%%%%%%%

In the present calculations, we use the following parameters: $g_\pi^2/4\pi=14.4$; another coupling constants from Table III of Ref.\ \cite{Ad04}; $f_\pi= 93$ MeV; a set of $\{\alpha_{00}^+=3.68, g_s=4.36, m_s=393 \mbox{~MeV}\}$ for the $(\pi,2\pi$) 3NF \cite{Ma00}; $\Lambda_\pi=800$ MeV and $\Lambda_\sigma=1300$ MeV. 
First, we note that the Argonne $V_{18}$ (AV18) 2NF  \cite{Wi95} underbinds the triton (${}^3$H) by 0.85 MeV, and the Brazil 2$\pi$E (BR${}_{800}$) 3NF \cite{Co83,Ro86} gives an  additional attraction of -1.75 MeV.
Thus a repulsive effect is expected to the $\pi$-S 3NF to complete the nuclear Hamiltonian.

Calculated values of the ${}^3$H energy for the AV18 plus each of the $\pi$-S 3NF models are presented in Table \ref{tab:pot_strength} as differences from the AV18 calculation.
This shows that 
a $\pi$-S 3NF with (positive) $V_a$-term produces an attractive contribution to the ${}^3$H energy, and that with (negative) $V_b$-term a repulsive contribution, which is consistent with the results given in Ref.\ \cite{Ad04}.
The $(\pi$-2$\pi)$ 3NF consists of  a negative $V_a$-term and a negative $V_b$-term, and produces a repulsive effect mostly due to the $V_a$-term.

In order to investigate effects of each term in the $\pi$-S 3NF on ND scattering observables, we pick up the following two $\pi$-S 3NF models to reproduce the ${}^3$H energy together with the AV18 2NF, the BR$_{800}$ 3NF, and a phenomenological spin-orbit type 3NF (SO) \cite{Ki99}:

$\cdot$ The $(\pi$-$\sigma)_{Z,PV}$ 3NF as a representative to $V_b$ term; 

$\cdot$ The $(\pi$-2$\pi)$ 3NF as a representative to $V_a$ term although it includes a small effect from $V_b$ term. 
(In this case, we take $\Lambda_\sigma=800$ MeV.)

The inclusion of the SO 3NF is effective to reproduce the VAPs, but gives only minor effects on the ${}^3$H energy and the TAPs at low energies.

\begin{figure}[tbh]
\begin{minipage}[t]{80mm}
\includegraphics[scale=0.65]{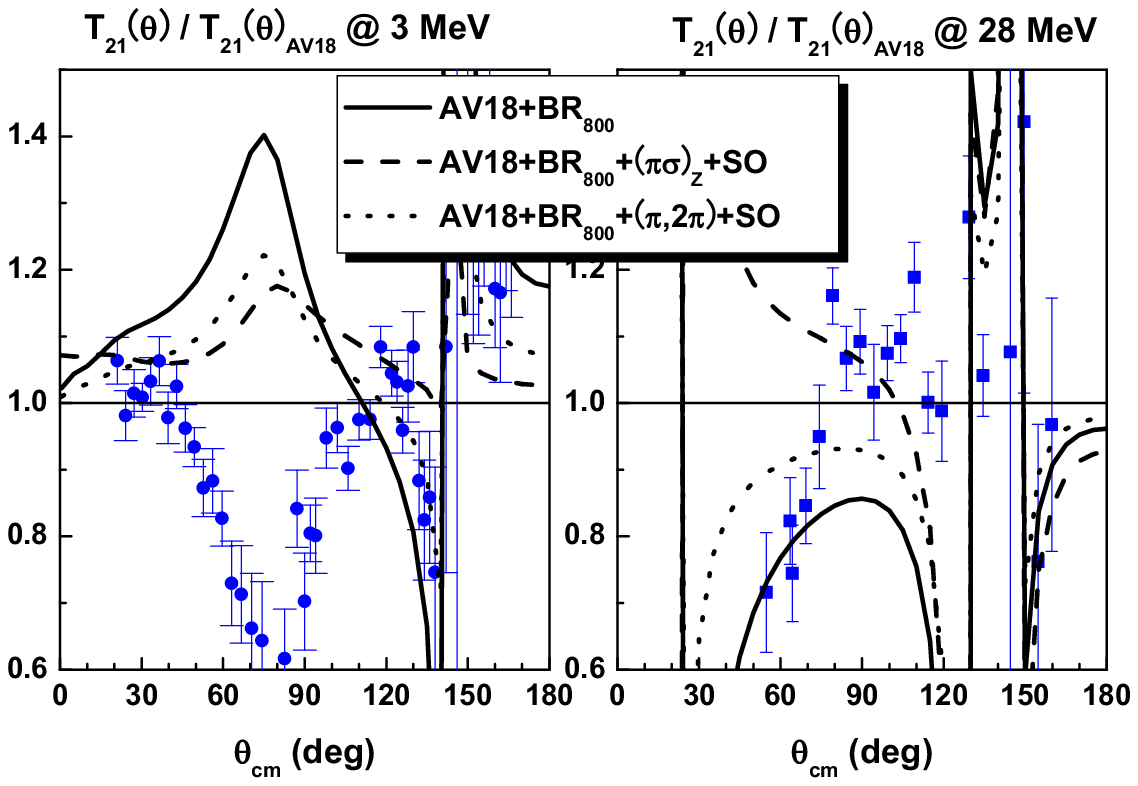}
\caption{ 
$T_{21}(\theta)$ of $pd$ elastic scattering normalized by the AV18 calculation at $E = 3$ MeV and $E = 28$ MeV.
The data are taken from Refs.\ \protect\cite{Sa94,Sh95} for 3 MeV and from Ref.\ \protect\cite{Ha84} for 28 MeV.
\label{fig:t21-ps-3-28MeV}
}
\end{minipage}
\hspace{\fill}
\begin{minipage}[t]{77mm}
\includegraphics[scale=0.65]{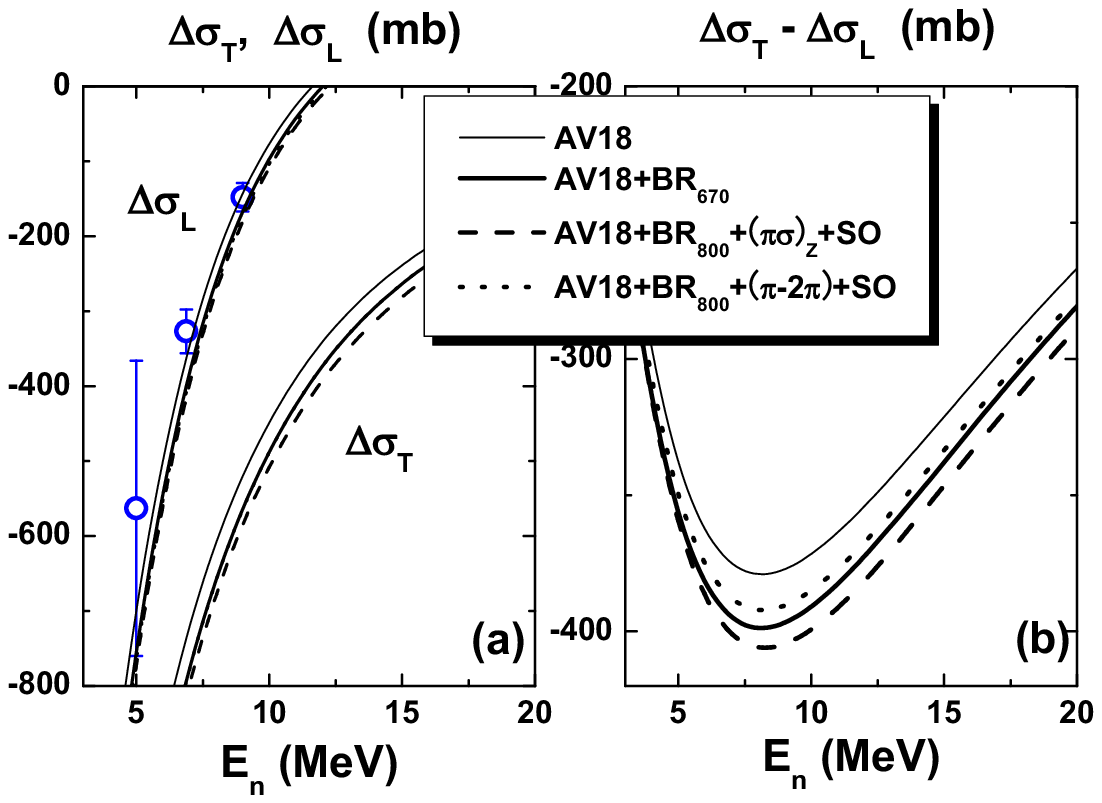}
\caption{ 
Spin-dependent total cross sections $\Delta\sigma_T$ and $\Delta\sigma_L$ (a), and their difference (b).
The data of $\Delta\sigma_L$ are taken from Ref.\ \protect\cite{Fo06}.
\label{fig:del-sig-tl-ps}
}
\end{minipage}
\end{figure}

Numerical results of Faddeev calculations \cite{Is06} for the tensor analyzing power $T_{21}(\theta)$ at $E=3$ MeV and $E=28$ MeV are presented in Fig. \ref{fig:t21-ps-3-28MeV}, where we plot the experimental data and the results with the 3NFs divided by the AV18 calculations.

For $T_{21}(\theta)$ at 3 MeV, both $\pi$-S 3NFs equally tend to cancel the effect due to the 2$\pi$E 3NF opposite to the data, but still leave an amount of discrepancy. 

At 28 MeV, the 3NFs contribute to $T_{21}(\theta)$ in different ways depending on scattering angles. 
At scattering angles of $50^{\circ}$ to $80^{\circ}$, the calculation with the BR${}_{800}$ 3NF and the one with the BR${}_{800}$ + $(\pi,2\pi)$ 3NFs look consistent with the data. 
On the other hand, at scattering angles of $90^{\circ}$ to $120^{\circ}$, 
the calculation with the BR${}_{800}$ + $(\pi$-$\sigma)_{Z,PV}$ 3NFs as well as the one with only 2NF look consistent with the data.

In Fig. \ref{fig:del-sig-tl-ps} (a), we display results of the transversal $\Delta\sigma_T$ and the longitudinal $\Delta\sigma_L$ asymmetries of the spin dependent total cross sections in $\vec{n}-\vec{d}$ scattering comparing with recent experimental data of $\Delta\sigma_L$ \cite{Fo06}. 
In Ref.\ \cite{Is03b}, we showed that effects of tensor interactions are prominent in the difference of $\Delta\sigma_T - \Delta\sigma_L$. 
As expected, differences of tendency in tensor components of the 3NFs are significantly observed in Fig. \ref{fig:del-sig-tl-ps} (b).

%----- FIGURE  -------------------------------------------------------
%----- FIGURE  -------------------------------------------------------

%%%%%%%%%%%%%%%%%%%%%%%%
\section{SUMMARY}
\label{sec:summary}
%%%%%%%%%%%%%%%%%%%%

We have examined three-nucleon forces arising from the exchange of $\pi$ and scalar-isoscalar object among three nucleons: 
$\pi$-$\sigma$ exchange via Z-diagram ($V_b$-term in Eq.\ (\ref{eq:Wps})); 
$\pi$-effective 2$\pi$ exchange ($V_a$-term, small $V_b$-term).
Both models produce repulsive effects on the ${}^3$H energy to compensate strong attraction caused by the 2$\pi$E 3NF.
Each 3NF affects polarization observables of nucleon-deuteron scattering, $T_{21}(\theta)$ and $\Delta\sigma_{T}-\Delta\sigma_{L}$, in different ways.
This shows that further measurements of these observables at some energies provide additional information on three-nucleon forces, which should be included in the nuclear Hamiltonian in addition to the 2$\pi$E 3NF.
%

%----- End of MAIN TEXT ----------------------------------------------

\newcommand{\etal}{{\em et al.}}

%----- REFERENCES  --------------------------------------------------

\end{document}